\documentclass[preprint, 
superscriptaddress,
 amsmath,amssymb,
prb,
floatfix,
]{revtex4-1}

\usepackage{graphicx}
\usepackage{dcolumn}
\usepackage{bm}
\usepackage{ulem}

\usepackage{color,soul}
\usepackage{epstopdf}
\AppendGraphicsExtensions{.tif}
\begin{document}


\title{Effect of strain on magnetic and orbital ordering of LaSrCrO$_3$/LaSrMnO$_3$ heterostructures }
\author{Sanaz Koohfar}
 \affiliation{Department of Physics, North Carolina State University, Raleigh, NC, 27695 USA}

\author{Alexandru B. Georgescu}
 \affiliation{Center for Computational Quantum Physics, Flatiron Institute, 162 5th Avenue, New York, NY, 10010, USA}

\author{Ingrid Hallsteinsen}
 \affiliation{Department of Electronic Systems, Norwegian University of Science and Technology, N-7491 Trondheim, Norway}
 \affiliation{Lawrence Berkeley National Laboratory, Berkeley, CA 94720, USA}

\author{Ritesh Sachan}
\affiliation{Mechanical and Aerospace Engineering, Oklahoma State University, Stillwater, OK 74078}

\author{Manuel A. Roldan}
\affiliation{Eyring Materials Center, Arizona State University, Arizona 85287, USA}
\author{Elke Arenholz}
 \affiliation{Lawrence Berkeley National Laboratory, Berkeley, CA, USA}
 \affiliation{Cornell High Energy Synchrotron Source, Ithaca, NY, USA}

\author{Divine P. Kumah}
\email{dpkumah@ncsu.edu}
 \affiliation{Department of Physics, North Carolina State University, Raleigh, NC, 27695 USA}

\date{\today}

\begin{abstract}
We investigate the effect of strain and film thickness on the orbital and magnetic properties of LaSrCrO$_3$ (LSCO)/LaSrMnO$_3$ (LSMO) heterostructures using bulk magnetometry, soft X-ray magnetic spectroscopy, first-principles density functional theory, high-resolution electron microscopy and X-ray diffraction. We observe an anti-parallel ordering of the magnetic moments between the ferromagnetic LSMO layers and the LSCO spacers leading to a strain-independent ferromagnetic ground state of the LSCO/LSMO heterostructures for LSMO layers as thin as 2 unit cells. As the LSMO thickness is increased, a net ferromagnetic state is maintained, however, the average magnetic moment per Mn is found to be dependent on the magnitude of the substrate-induced strain. The differences in the magnetic responses are related to preferential occupation of the Mn $x^2-y^2$ (in-plane) d-orbitals for tensile strain and $3z^2-r^2$ (out-of-plane) orbitals under compressive strain leading to competing ferromagnetic and anti-ferromagnetic exchange interactions within the LSMO layers. These results underscore the relative contributions of orbital, structural and spin degree of freedom and their tunability in atomically-thin crystalline complex oxide layers.


\end{abstract}

\maketitle

\section{Introduction}
The ability to achieve atomically-precise perovskite transition metal oxide interfaces using thin-film deposition techniques such as molecular beam epitaxy allows for the unprecedented control of spin, orbital and structural degrees of freedom.\cite{Tokura2000OrbitalOxides, brinkman2007magnetic,Hwang2012EmergentInterfaces, pesquera2016strain} In the transition metal oxides, the ordering and relative electronic occupation of the transition metal \textit{d} orbitals has important implications for the macroscopic functional properties of these systems including magnetism and superconductivity.\cite{aruta2009orbital,Chakhalian2012WhitherInterface,Hellman2017Interface-inducedMagnetism, Disa2015OrbitalHeterostructures,li2019superconductivity, sadoc2010large}  Decoupling the role of intrinsic interfacial interactions and strain on spin and orbital ordering at interfaces remains an active area of scientific research particularly, when film thicknesses are on the order of a few atomic layers.\cite{freeland2011orbital, Zubko2011InterfaceHeterostructures, Boris2011DimensionalitySuperlattices,Benckiser2011OrbitalHeterostructures, Gibert2012ExchangeSuperlattices} Rare-earth manganite-based heterostructures provide an ideal platform for the systematic investigation and control of these interactions by quantum confinement and epitaxial strain.\cite{Bhattacharya2014MagneticHeterostructures, takamura2008control, yang2010strain, colizzi2008interplay, Moon2014EffectFilms} However, for thin manganite films, a challenge remains in suppressing interface and surface polarity-driven chemical and structural distortions which suppress magnetic ordering and metallicity below a substrate strain-dependent critical thickness.\cite{Koohfar2017StructuralInterfaces, Liao2015Origin3, Herger2008StructureMagnetoresistance, Sun1999Thickness-dependentFilms, Borges2001MagneticFilms, Huijben2008CriticalFilms, Bhattacharya2014MagneticHeterostructures} For single layer La$_{0.7}$Sr$_{0.3}$MnO$_3$(LSMO) films, the magnetic critical thickness is reported to vary from 3 nm on NdGaO$_3$ (110) to 1.2-3 nm on SrTiO$_3$ (001) and 5 nm on LaAlO$_3$ (001).\cite{Huijben2008CriticalFilms, Sun1999Thickness-dependentFilms,Angeloni2004SuppressionFilms,Borges2001MagneticFilms, Kourkoutis2010MicroscopicLayers} The modification of the LSMO interface by the addition of spacer layers to couple to oxygen octahedral rotations and compensate the polar discontinuity present at LSMO interfaces has resulted in improved magnetization and electronic properties, however, a systematic demonstration of the removal of magnetic dead layers independent of the substrate-induced strain remains.\cite{Guo2018RemovalDesign, Peng2014TuningEngineering, Moon2017StructuralHeterostructures, boschker2012preventing}

The insertion of ultrathin La$_{0.7}$Sr$_{0.3}$CrO$_3$ (LSCO) layers at the interfaces of LSMO films grown on (001)-oriented SrTiO$_3$(STO) has been recently found to  result in the stabilization of ferromagnetism in LSMO layers as thin as 0.8 nm (2 unit cells (uc)).\cite{Koohfar2019ConfinementHeterostructures} At these length scales, intrinsic interfacial interactions are expected to dominate the physical properties of the oxide films. Jahn-Teller-like distortions induced by epitaxial strain breaks the degeneracy of the transition metal \textit{d}-orbitals via the crystal field splitting leading to changes in the electronic occupation of orbitals with in-plane and out-of-plane symmetries.\cite{Tebano2010PreferentialPhotoemission, Pesquera2012SurfaceFilms, Tokura2000OrbitalOxides, Millis2011OxideMagnetism} For unstrained LSMO, the double degeneracy of the Mn \textit{3d e$_g$} orbitals with out-of-plane $3z^2-r^2$ and in-plane $x^2-y^2$ symmetry is related to  double exchange interactions and a ferromagnetic (FM) metallic ground state. For compressive and tensile biaxial strained LSMO, C-type and A-type anti-ferromagnetic (AF) ground states are stabilized, respectively.\cite{konishi1999orbital, fang2000phase, colizzi2008interplay, Tebano2008EvidenceFilms, Pesquera2012SurfaceFilms, Aruta2006StrainFilms, Tebano2010PreferentialPhotoemission} In addition to the strain-dependent orbital ordering, at LSMO surfaces, a strain-independent preferential occupation of the $3z^2-r^2$ orbital \cite{Pesquera2012SurfaceFilms, Tebano2010PreferentialPhotoemission} has been observed related to inversion symmetry breaking and a surface structural reconstruction.\cite{Herger2008StructureMagnetoresistance, Koohfar2017StructuralInterfaces, Li2017ImpactHeterostructures, Gray2011InsulatingPhotoemission} We show that in LSCO/LSMO heterostructures, a net ferromagnetic ground state is stabilized independent of the substrate-induced strain state and LSMO thickness. In the ultra-thin limit, for 2 uc of LSMO, the dominant physics is set by the interfacial spin interactions, while with increasing LSMO thickness, strain-induced orbital ordering dictates the physical properties of the heterostructures. 

In the current work, we investigate the effect of strain on orbital and magnetic ordering in [2 uc LSCO/ N uc LSMO] superlattices, where N, the LSMO thickness is varied from 2 to 8 unit cells (uc) for heterostructures grown on (001)-oriented LaAlO$_3$ (LAO),  (La$_{0.18}$Sr$_{0.82}$)(Al$_{0.59}$Ta$_{0.41}$)O$_3$ (LSAT) and SrTiO$_3$ (STO) by molecular beam epitaxy (MBE). The structural, orbital and magnetic properties of [2 LSCO/ N LSMO] heterostructures are investigated by temperature-dependent SQUID measurements, X-ray diffraction (XRD), X-ray linear dichroism (XLD) and X-ray magnetic circular dichroism (XMCD) and first principles density functional theory. XLD measurements indicate a transition from $x^2-y^2$ in-plane orbital ordering for tensile strained layers to $3z^2-r^2$ out-of-plane ordering for compressively strained films. For the thinnest LSMO layers (2 uc), the interfacial spin and structural interactions between the LSMO and LSCO layers leads to a net FM ground state characterized by an AF coupling of Mn and Cr spins across the interface.\cite{Koohfar2019ConfinementHeterostructures, ziese2010tailoring} LSCO in bulk is a G-type antiferromagnet,\cite{tezuka1998magnetic} however, the interfacial coupling with LSMO leads to a net anti-alignment of the Cr spins with the applied magnetic field. As the LSMO thickness is increased, a net FM state is still maintained, however the effects of epitaxial strain and orbital ordering are non-negligible leading to competing FM and AF exchange interactions within the LSMO layers. 

\section{Results}
\subsection{Growth}
The MBE technique was used to to grow [ M uc LSCO/ N uc LSMO]$_x$, (where \textit{x} is the number of bilayer repeats) superlattices (SLs) on single crystalline (001) oriented LSAT, STO, and LAO substrates. M was fixed at 2 uc and N was varied from 2 to 8 ucs. The [2 LSCO/ 2 LSMO]$_{10}$, [2 LSCO/ 4 LSMO]$_{6}$, [2 LSCO/ 6 LSMO]$_{4}$ and [2 LSCO/ 8 LSMO]$_{4}$ SLs were capped with 2 ucs of LSCO. The La, Sr, Cr and Mn fluxes were calibrated prior to growth with a quartz crystal microbalance. After growth at 900 $^o$C, the heterostructures were cooled down to room temperature in 7x10$^{-6}$ Torr oxygen plasma. \textit{In-situ} reflection high energy electron diffraction was used to monitor the film crystallinity and thickness during growth. 

\subsection{X-ray Diffraction Measurements}
\begin{figure*}[ht]
\centering
\includegraphics[width=0.85\textwidth]
{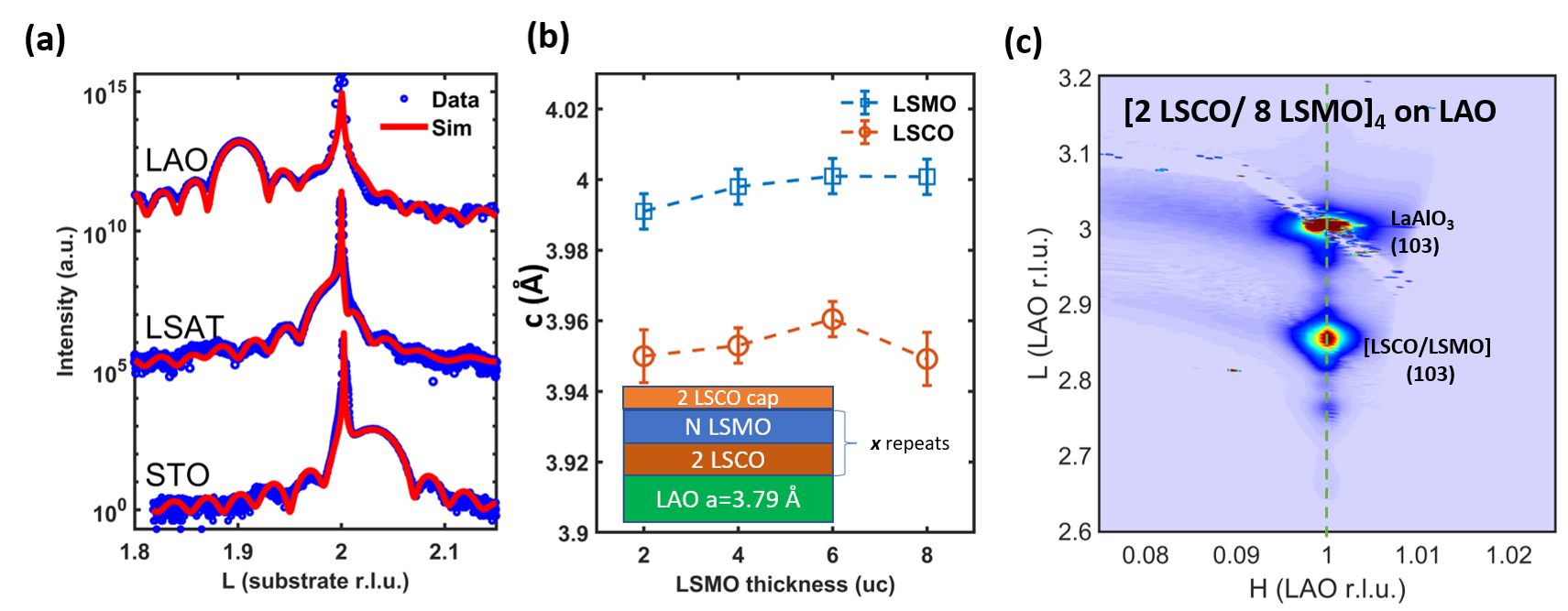}
\caption{\textbf{X-ray diffraction measurements for [2 LSCO/ N LSMO]$_x$ as a function of epitaxial strain}. (a) Measured specular X-ray diffraction and fits for [2 LSCO/ 6 LSMO]$_{4}$ superlattices grown epitaxially on LAO, LSAT and STO. $L$ are in reciprocal lattice units of the respective substrates (1 $r.l.u. = 1/a_{substrate} (\AA^{-1})$). (b) Measured out-of-plane lattice constants, \textit{c}, for LSMO and LSCO layers in [2 LSCO/ N LSMO]$_x$ on (001) LAO as a function of N, the LSMO thickness. (c) Reciprocal space map around the (103) Bragg peak for a [2 LSCO/ 8 LSMO]$_{4}$ SL on LAO.}
\label{fig:XRD}
\end{figure*}

X-ray diffraction measurements around the film and substrate (002) Bragg peaks were performed to determine the lattice constants and strain states of the heterostructures. Figure \ref{fig:XRD}(a) shows representative scans for [2 LSCO/ 6 LSMO]$_{4}$ superlattices grown on LAO (a=0.379 nm), LSAT (a=0.3868 nm) and STO (a=0.3905 nm). The presence of finite thickness oscillations indicate high crystallinity and sharp interfaces which are confirmed by high resolution electron microscopy measurements and electron energy loss spectroscopy measurements.

The out-of-plane lattice constants for the LSMO and LSCO layers in the [2 LSCO/ 6 LSMO]$_{4}$ superlattices were determined from fits to the measured diffraction data to be 0.4004 nm (LAO), 0.3891 nm (LSAT) and 0.3846 nm (STO). The fits were performed by varying the LSMO and LSCO lattice constants using the GenX software.\cite{Bjorck2007GenX:Evolution} Bulk LSMO has a pseudocubic lattice constant of 0.3875 nm,\cite{lebedev2003structure} hence the measured lattice constants are indicative of coherently strained films with 2\% and 0.2\% compressive lattice mismatch on LAO and LSAT respectively, and 1\% tensile mismatch on STO. The lattice mismatch, $\delta$ is determined from the bulk pseudocubic lattice constant by the relation  $$\delta =\frac{a_{bulk LSMO}-a_{substrate}}{a_{bulk LSMO}}$$.

Figure \ref{fig:XRD}(b) shows the measured LSMO lattice constant for [2  LSCO/ N LSMO] for N=2,4, 6 and 8 ucs for SLs grown on LAO indicating that the LSMO and LSCO layers remain fully compressively strained up to N=8 uc. The increase in the LSMO lattice constant with increasing LSMO thickness is attributed to a slight increase in oxygen vacancies.\cite{orgiani2012evidence} A post-growth anneal leads to a contraction of the c-lattice parameter of the [2  LSCO/ 4 LSMO]$_6$ by 0.01 \AA{}, however, SQUID measurements indicate a minimal change in the magnetic properties of the films (Figure S1).\cite{supplement}  To confirm that the films are coherently strained, reciprocal space maps were measured around the film and substrate (1 0 3) Bragg peaks at beamline 33ID at the Advanced Photon Source. A representative RSM for a [2 LSCO/ 8 LSMO]$_4$ on LAO is shown in Figure \ref{fig:XRD}(c) confirms that the in-plane lattice constant of the film on LAO is 0.379 \AA{}. 

\subsection{Transmission electron microscopy}

To gain further insight into the atomic-scale structure of the heterostructures, atomic resolution electron microscopic studies were conducted. Aberration-corrected STEM-JEOL ARM was used in conjunction with EELS to acquire HAADF images and EEL spectra of the superlattice structures. In the experiment, the microscope was operated at an accelerating voltage of 200 kV and the electron probe current of 38 $\pm$ 2 pA. The convergence and collection angles in the experiments were 19 and 65 mrad, respectively. The EELS data were acquired with a collection angle of 28 mrads.

\begin{figure*}[ht]
\centering
\includegraphics[width=0.85\textwidth]
{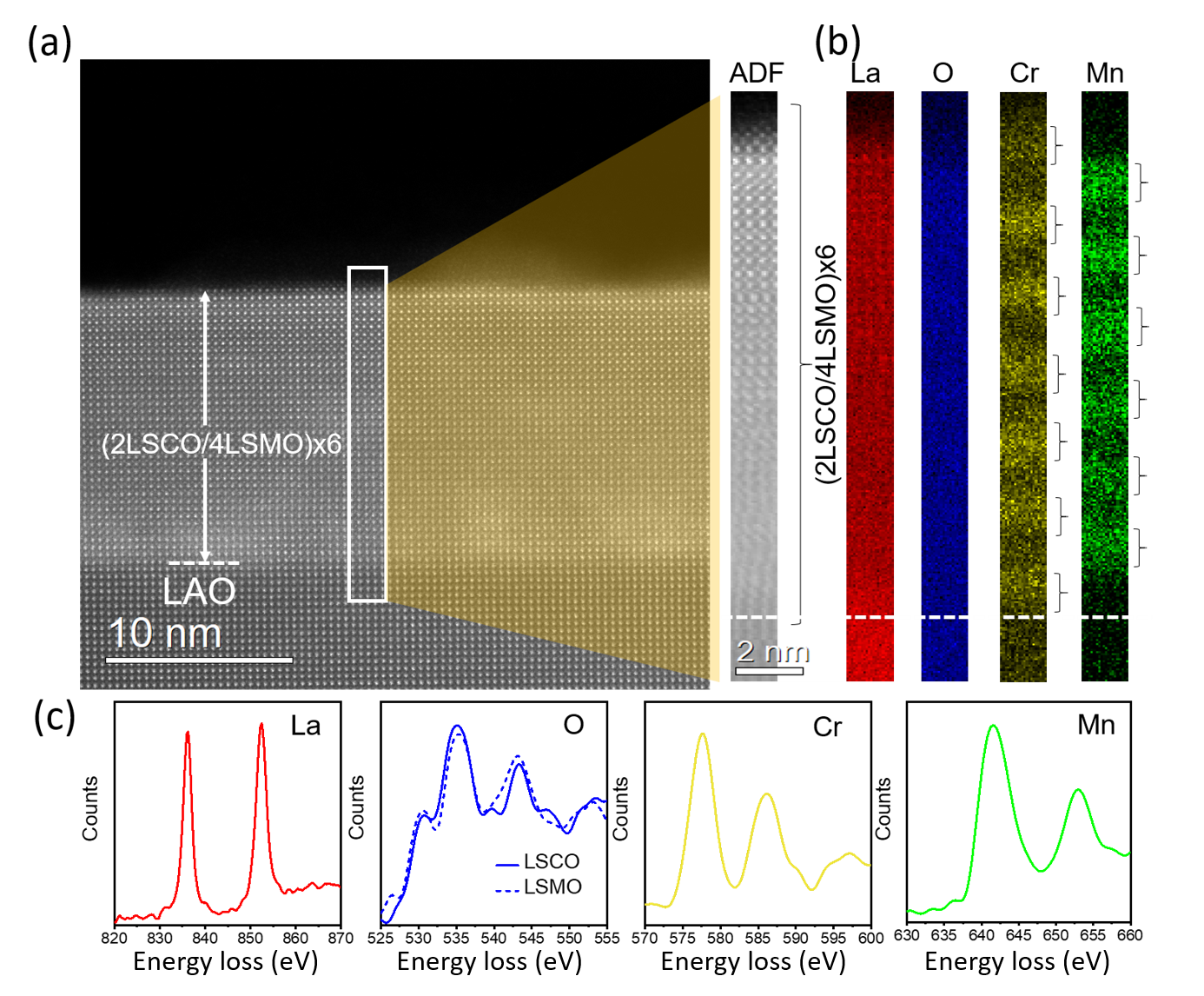}
\caption{\textbf{High-resolution transmission electron microscope images of a [2 LSCO /4 LSMO]$_6$ grown on (001)-LaAlO$_3$}. (a) High-angle angular dark field image along the LAO (100) zone axis (b) elemental maps obtained by electron energy loss spectroscopy measurements (c) representative EELS spectra for La, O, Cr and Mn. EELS spectra are shown for O within the LSMO and LSCO sub-layers.}
\label{fig:TEM}
\end{figure*}

A representative cross-section high-angle annular dark field (HAADF) image of a [2 LSCO/ 4 LSMO]$_{6}$ SL structure on LAO is shown in Figure \ref{fig:TEM}(a), showing high-quality, defect-free epitaxial growth of the superlattice on the substrate with an atomically abrupt interface between the LAO substrate and the first LSCO layer. Due to the similarities in electron densities of the Cr and Mn ions, the LSCO and LSMO interfaces are not directly distinguishable in the HAADF image. Thus, electron energy loss spectroscopy (EELS) measurements were performed to distinguish the LSCO and LSMO layers.
The elemental maps determined from representative La-$M_{5,4}$, O-\textit{K}, Cr-$L_{3,2}$ and Mn-$L_{3,2}$ EELS-edges from the superlattice are shown in Figure \ref{fig:TEM}(b). The elemental maps clearly indicate the synthesis of well-defined LSCO and LSMO layers and a uniform distribution of La and O throughout the heterostructure. 
 From the Cr and Mn elemental maps, minor intermixing between LSMO/LSCO interface is observed, which could possibly be due to the EELS signal delocalization within the monolayers in addition to the partial abruptness at the interface.\cite{Koohfar2019ConfinementHeterostructures} Figure \ref{fig:TEM}(c) shows representative EELS spectra for La, O, Cr and Mn respectively. For the O-K edge spectra, a pronounced prepeak at 526 eV is observed for O in the LSMO layers indicative of a stronger Mn-O orbital hybridization than the Cr-O hybridization.

\subsection{Magnetization}
\begin{figure*}[ht]
\centering
\includegraphics[width=0.85\textwidth]{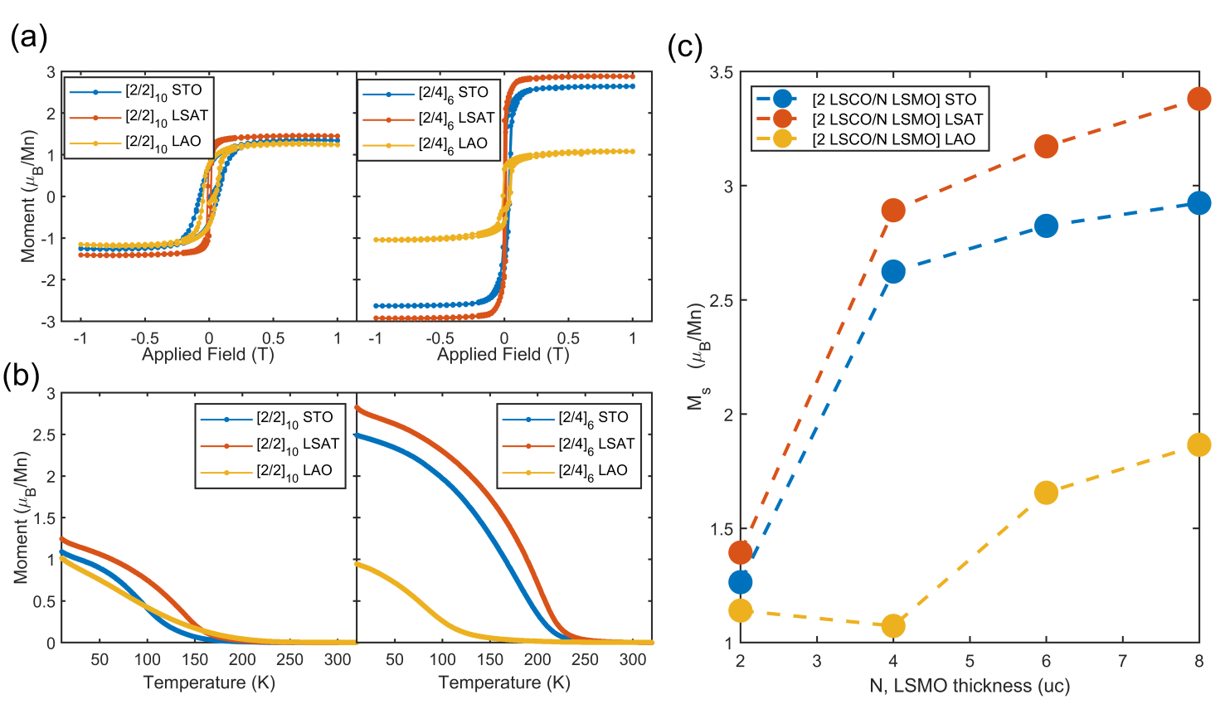}
\caption{\textbf{Magnetic properties of [2 LSCO/ N LSMO]$_x$ heterostructures as a function of strain and LSMO thickness.} Magnetization for [2 LSCO/ 2 LSMO]$_{10}$ and [2 LSCO/ 4 LSMO]$_6$ superlattices grown on (001)-oriented STO, LSAT and LAO as a function of (a) in-plane magnetic field at 10 K and (b) temperature. (c) Saturation magnetization, $M_{s}$, from magnetization versus applied magnetic field at 10 K as a function of LSMO thickness, N, and strain at 10 K for [2uc LSCO/ N uc LSMO]$_x$ superlattices. }
\label{fig:SQUID}
\end{figure*}

The temperature-dependent magnetization of the [LSCO/LSMO] heterostructures was determined as a function of strain and LSMO thickness by SQUID  measurements from 5-350 K with a magnetic field applied along the LSMO in-plane [100] axis. The magnetization as a function of temperature curves were measured on warming with a 0.1 T field applied in-plane after a field cool in a 0.2 T magnetic field. Figure \ref{fig:SQUID}(a) and \ref{fig:SQUID}(b) show the magnetization as a function of applied magnetic field at 10 K and temperature, respectively, for [2 LSCO/ 2 LSMO]$_{10}$ and [2 LSCO/ 4 LSMO]$_{6}$ SLs grown on LAO, STO and LSAT. 

The saturation magnetization M$_s$ obtained from the magnetization versus applied magnetic field measured at 10 K is plotted in Figure \ref{fig:SQUID}(c) as a function of strain and the LSMO thickness. For each LSMO thickness, M$_s$ is found to be inversely proportional to the magnitude of the substrate-induced strain. 


 For the relatively unstrained heterostructure on LSAT (c/a=1.006), the magnetization increases from 1.3 $\mu_B/Mn$ for N=2 to a value of 3.4 $\mu_B/Mn$ for N=8 close to the expected magnetic moment for bulk LSMO. The increase in the SQUID magnetization with LSMO thickness is related to the reduced contribution to the total magnetization from the LSCO layers with moments anti-aligned to the magnetic field.\cite{Koohfar2019ConfinementHeterostructures} A similar increase is observed for the heterostructures on STO. For the LAO strain, the net moment increases slightly from 1.1 $\mu_B/Mn$ for N=2 to 1.8 $\mu_B/Mn$ for N=8. Additionally, the Curie temperature increases with the net M$_s$ of the samples as shown in Figure \ref{fig:SQUID}(b) and Figure S2(a) of the Supplementary materials\cite{supplement} where the magnetization as a function of temperature are compared for the SLs on LSAT. The Curie temperature and M$_s$ scale with the LSMO thickness and the LSMO/LSCO thickness ratio (Figure S2(b) of Supplementary Materials) \cite{Koohfar2019ConfinementHeterostructures} indicating that the energy scales for FM ordering are set by the LSMO/LSCO interfacial magnetic interactions.

\subsection{X-ray magnetic circular dichroism measurements}
To determine the element-specific magnetic ordering in the LSMO and LSCO layers as a function of strain, X-ray magnetic circular dichroism (XMCD) measurements were performed at the Mn and Cr L-edges at the 4.0.2 magnetic spectroscopy beamline at the Advanced Light Source. The measurements were performed in total electron yield (TEY) mode at 80 K with an 0.5 T magnetic field applied parallel to the film surface and a photon grazing angle of incidence of 30$^\circ$. The XMCD signal is determined from the differences between the  XAS measured with right and left circularly polarized light.

\begin{figure*}[ht]
\centering
\includegraphics[width=0.85\textwidth]{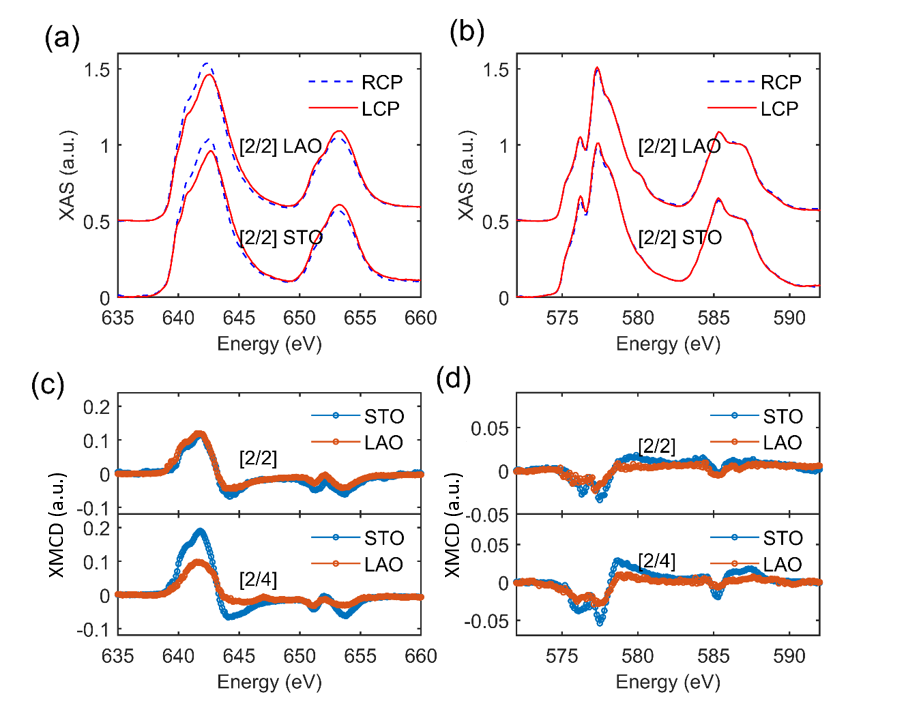}
\caption{\textbf{Element-specific magnetic measurements.} Comparison of X-ray absorption spectra measurements with right( RCP) and left (LCP) circularly polarized light at 80 K at (a) the Mn L-edge  and (b) the Cr L-edge for [2 LSCO/2 LSMO]$_{10}$ heterostructures grown on STO and LAO. Comparison of XMCD measurements for [2 LSCO/2 LSMO] and [2 LSCO/4 LSMO]$_6$ SLs on STO and LAO at the (c) Mn L-edge and (d) Cr L-edge.}
\label{fig:XMCD}
\end{figure*}

A comparison of the polarization-dependent XAS signal at the Mn L-edge is shown in Figure \ref{fig:XMCD}(a) for the [2 LSCO/ 2 LSMO]$_{10}$ SLs on LAO and STO at 80 K. XAS measurements at the Cr L-edge are shown in Figure \ref{fig:XMCD}(b). The XMCD for the [2 LSCO/ 2 LSMO]$_{10}$ and [2 LSCO/ 4 LSMO]$_{6}$ SLs on STO and LAO are compared for the Mn L-edge and the Cr L-edge in Figure \ref{fig:XMCD}(c) and \ref{fig:XMCD}(d) respectively.

For all samples, the integrated XMCD signal at the Mn L$_3$-edge (from 635 eV to 648 eV) are positive, indicative of  net ferromagnetic ordering in the LSMO layers. In agreement with previous reports on [LSCO/LSMO] heterostructures grown on STO,\cite{Koohfar2019ConfinementHeterostructures}  a positive Mn XMCD signal is observed for all the [LSCO/LSMO] heterostructures on LAO, STO and LSAT for LSMO thicknesses as thin as 2 ucs confirming that the presence of the LSCO layers results in the removal of magnetic dead LSMO layers and the stabilization of a FM LSMO ground state.
The integrated Cr L$_3$ XMCD spectra in Figure \ref{fig:XMCD}(d) are negative, indicating a net alignment of the moments on the Cr sites anti-parallel to the applied magnetic field. The relatively weak Cr signal may be attributed to the sensitivity of the total electron yield to the surface layers. The films are capped with LSCO and previous XRD measurements indicate structural distortions of the surface layers.\cite{Koohfar2019ConfinementHeterostructures} 
The anti-parallel configuration of the spins in the LSMO and LSCO layers is attributed to an AF superexchange between the interfacial Mn and Cr.\cite{Koohfar2019ConfinementHeterostructures} Here, we find that the interfacial AF exchange and the FM ground state of the [LSCO/LSMO] heterostructure is independent of epitaxial strain.


As the LSMO thickness is increased to 4 uc in the [2 LSCO/ 4 LSMO]$_{6}$ SLs, the XMCD signal shown in Figure \ref{fig:XMCD}(c) at the Mn L-edge increases for the SL on STO relative to the SL on LAO. XMCD sum rules\cite{stohr1995determination, Yi2016Atomic} are applied to the XMCD data to quantitatively compare the element-specific magnetic moments. The XMCD-derived Mn spin (m$_s$) and orbital(m$_l$) magnetic moments are compared in Table S1 of the supplemental materials for the [2 LSCO/ 2 LSMO] and [2 LSCO/ 4 LSMO] SLs on LAO and STO.\cite{supplement} While the absolute values of the XMCD-derived magnetic moments may be lower than the moments predicted by SQUID due to the finite probe depth (~3 nm) of the TEY mode,\cite{nakajima1999electron} the assumption that the magnetic dipole term is negligible \cite{Shibata2017} and errors due to the overlap of the L$_2$ and L$_3$ peaks, the trends for samples with similar thicknesses can be compared. For the [2 LSCO/ 2 LSMO] SLs, m$_s$ is 1.07 and 1.23 $\mu_B$/Mn for LAO and STO respectively. For the [2 LSCO/ 4 LSMO] SLs, m$_s$ is 0.99 and 1.64 $\mu_B$/Mn for on LAO and STO respectively. The increase in m$_s$ on STO is in agreement with the SQUID measurements in Figure \ref{fig:SQUID}. 


\subsection{X-ray linear dichroism measurements}
\begin{figure*}[ht]
\centering

\includegraphics[width=0.9\textwidth]{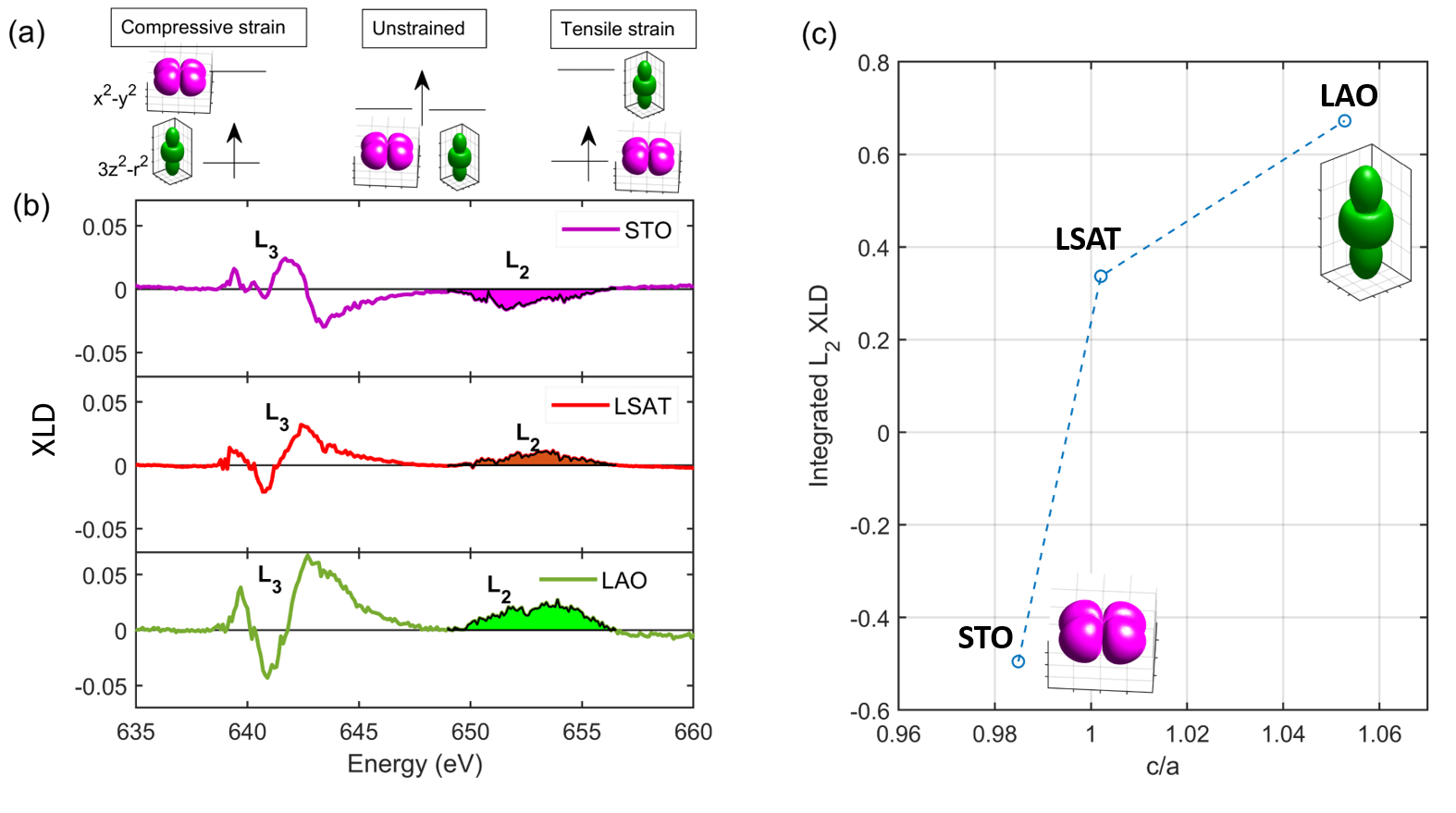}
\caption{\textbf{Orbital ordering in [2 LSCO/ N LSMO]$_x$ probed by X-ray linear dichroism.} (a) Schematic of predicted ordering of the Mn \textit{$e_g$} orbitals as a function of epitaxial strain. (b) X-ray linear dichroism measurements at the Mn L-edge for [2 LSCO/ N LSMO]$_x$ superlattices grown on STO, LSAT and LAO. Measurements were performed at 300 K. The XLD spectra are determined from $I_\parallel- I_\perp$, where $I_\parallel$ and  $I_\perp$ are the absorption spectra measured with the incident X-ray polarization oriented parallel to the film surface and perpendicular to the film surface respectively. N=4 on STO and LAO and N=6 on LSAT. (c) Plot of the area under the Mn L$_2$ XLD peaks shaded in (b) as a function of the LSMO c/a ratio.}
\label{fig:LD}
\end{figure*}

To determine the orbital polarization of the films as a function of strain, XLD measurements were carried out at Mn L$_{2,3}$-edges to determine the symmetry of unoccupied Mn \textit{3d e$_g$} states. The XLD is determined from the difference between the X-ray absorption spectra (XAS) measured with the photon polarization perpendicular (I$_\perp$) and parallel (I$_\parallel$)  to the film surface, \textit{i.e.} $XLD=I_\parallel-I_\perp$. Since, the XAS spectra probe the density of unoccupied Mn 3\textit{d} states,  a positive (negative) XLD is indicative of a preferential occupation of the out-of-plane 3z$^2-r^2$ (in plane, $x^2-y^2$)  orbitals. A schematic of the strain-induced ordering of the Mn 3d e$_g$ orbitals is shown in Figure \ref{fig:LD}(a). The measured XLD spectra for [LSMO/LSCO] SLs on STO, LSAT and LAO are shown in Figure \ref{fig:LD}(b). A clear change in the sign of the XLD signal for the Mn $L_2$ peak is observed from negative to positive as the substrate-induced strain is varied from tensile on STO to compressive strain on LAO. The line shapes are consistent with previous measurements and theoretical calculations for strained LSMO films.\cite{Tebano2008EvidenceFilms, Aruta2006StrainFilms, Li2016CompetingHeterostructures, li2017origin}

The orbital polarization is quantified by plotting the area of the shaded regions in Figure \ref{fig:LD}(b) as a function of the c/a ratio in Figure \ref{fig:LD}(c). The change in the integrated L$_2$ XLD sign from negative for the tensile strained sample on STO to positive for the compressively strained sample on LSAT and LAO is consistent with the predicted change in orbital occupation due the strain-induced Jahn-Teller-like distortion.\cite{Tokura2000OrbitalOxides} Additionally, the magnitude of the XLD signal is proportional to the magnitude of the substrate-induced strain.



\subsection{Charge Transfer}
\begin{figure*}[ht]
\includegraphics[width=0.8\textwidth]{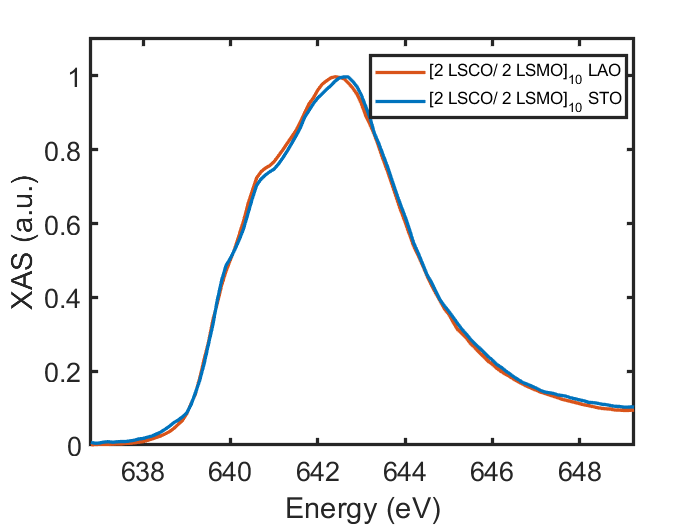}
\caption{\textbf{X-ray absorption spectroscopy measurements} for [2 LSCO/ 2 LSMO]$_{10}$ superlattices grown on LaAlO$_3$ and SrTiO$_3$ at the Mn L$_2$ edge. }
\label{fig:LAOXAS}
\end{figure*}

For the [2 LSCO/ 2 LSMO] SLs where interfacial interactions dominate the magnetic properties, we investigate the possibility of charge transfer between the LSMO and LSCO layers by comparing the Cr and Mn spectra as a function of strain. X-ray absorption spectra at 300 K are shown in Figure \ref{fig:LAOXAS} for [2 LSCO/ 2 LSMO]$_10$ on LAO and STO. As the strain changes from 1\% tensile strain on STO to 2\% compressive strain on LAO, two features of the Mn L$_2$ peak are observed. There is a shift to lower energies and an increase intensity of the lower energy shoulder at 641 eV. These features are consistent with an increase in the Mn$^{3+}$ on LAO compared to STO signifying electron transfer from Cr to Mn.\cite{lee2010hidden} Based on measured Mn white line shifts \cite{Tan2012OxidationEELS}, we estimate \~0.1 more holes (electrons) in the LSCO (LSMO) layers on LAO compared to STO.This is consistent with the slightly reduced magnetization of the [2 LSCO/ 2 LSMO]$_{10}$ samples on LAO measured by SQUID in Figure \ref{fig:SQUID}(c) and first principles calculations described below.


\subsection{Theory}
To confirm the effect of strain on the magnetization, first principles density functional theory calculations were performed for [2 LSCO/2 LSMO] SLs and bulk LSMO and LSCO with the in-plane lattice constants set to bulk LAO to simulate compressive strain. Our calculations were performed using the PBE exchange-correlation functional and ultrasoft pseudopotentials.\cite{Giannozzi2009QUANTUMMaterials, Perdew1996GeneralizedSimple, Vanderbilt1990SoftFormalism} A k-mesh of 7x7x2 was used for the [2 LSCO/2 LSMO] heterostructure calculations and 7x7x5 for the 20 atom bulk unit cell calculations with an energy cutoff of 50.0 Rydberg and a density cutoff of 400.0 Rydberg. The electronic calculations were converged to $10^{-9}$ Rydberg.

\begin{figure*}[ht]
\includegraphics[width=0.8\textwidth]{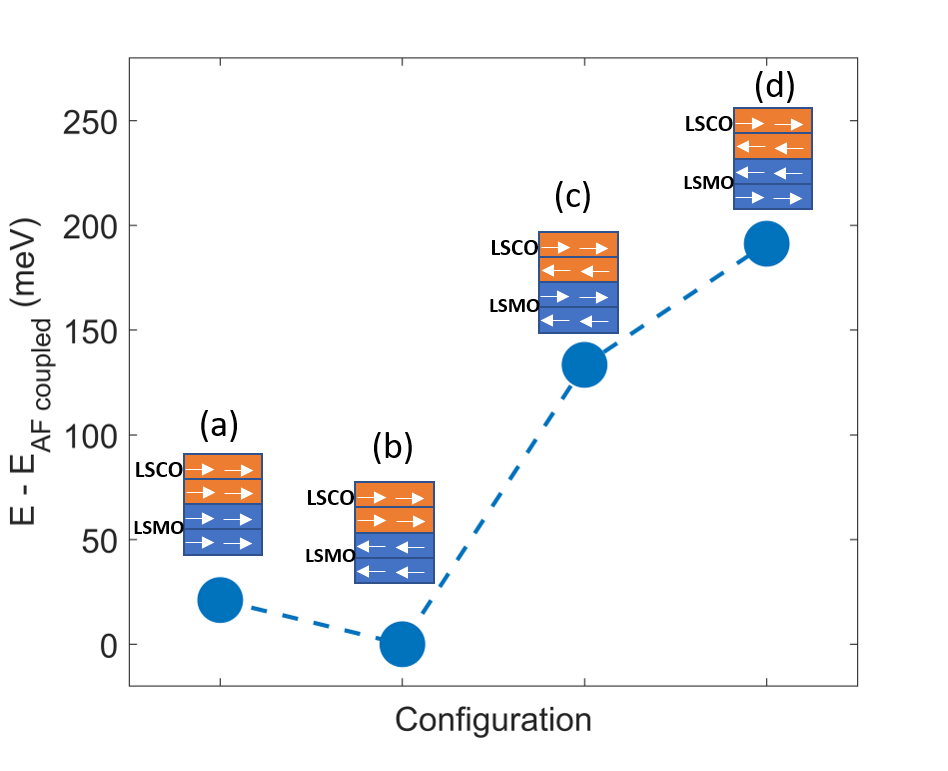}
\caption{\textbf{Calculated energies for magnetic configurations theoretically investigated for [2 LSCO/2 LSMO] heterostructures compressively strained to LaAlO$_3$}. The minimum energy state, E$_{AF coupled}$, corresponds to configuration (b) where the spins are aligned within each layer but are anti-aligned between the layers. The energies of the states are compared with the minimum energy state.}
\label{fig:DFT}
\end{figure*}

Our calculations for fully relaxed unstrained bulk LSMO and LSCO predict a FM ground state  AFM-C  magnetic structure respectively. To study the interlayer magnetic couplings between the LSCO and LSMO layers in a [2 LSCO/2 LSMO] heterostructure, we compare the total energies of different spin configurations as shown in Figure \ref{fig:DFT}. We find that the ground state for the heterostructure compressively strained to LAO is the AF configuration that presents spins aligned within each material, and anti-aligned across the interface, with an energy lower by 21 meV than the fully FM state, while the other two antiferromagnetic states that would have an effective total moment of 0 have significantly higher total energies (112 meV and 170 meV higher, respectively, than the state with AF-coupled FM layers). 

For the [2 LSCO/2 LSMO] SL compressively strained to LAO, the calculated magnetic moment per Mn for the AF-coupled ground state is 3.66 $\mu_B$/Mn in the LSMO layers and -2.65 $\mu_B$/ Cr in the LSCO layers. Previous calculations for a [2 LSCO /2 LSMO] SL under tensile strain on STO yield moments of 3.48 $\mu_B$/Mn and -2.8 $\mu_B$/Cr.\cite{Koohfar2019ConfinementHeterostructures} The decrease (increase) in the magnitude of the Cr(Mn) magnetic moments as the strain is varied from tensile to compressive strain suggests increased electron transfer from Cr to Mn on LAO. A displacement of the interfacial Oxygen atom of 0.024$\AA$ towards the Cr for the [2 LSCO/2 LSMO] superlattice on LAO provides further indication of a minor amount of electron transfer from the Cr to Mn, favoring the Cr$^{4+}$ and Mn$^{3+}$ configurations, respectively.

\section{Discussion}
The SQUID measurements in Figure \ref{fig:SQUID} and the XMCD measurements in Figure \ref{fig:XMCD} confirm FM ordering within the LSMO layers in the [LSCO/LSMO] heterostructures independent of the applied strain. For the [2 LSCO/2 LSMO]$_{10}$ samples, the comparable SQUID magnetization is consistent with AF interactions across the LSCO/LSMO interface as the dominant mechanism leading to the FM in the LSMO layers. Following reference \cite{Koohfar2019ConfinementHeterostructures}, the Mn and Cr moments are estimated to be 3.4 $\mu_B$/Mn and -2.1 $\mu_B$/Cr and the total SQUID magnetization per Mn is predicted to be 1.3-1.4$\mu_B$/Mn. This prediction is consistent with the measured magnetization in Figure \ref{fig:SQUID} for the [2 LSCO/ 2 LSMO]$_{10}$ samples. The AF coupling between the LSCO and LSMO layers is confirmed by exchange bias measurements in Figure S3 of the Supplementary Materials.\cite{supplement}

To understand the strain-independent interfacial AF coupling between the Cr and Mn, we consider the Goodenough-Kanamori rules for predicting exchange interactions between neighbouring empty and half-filled transition metal \textit{d}-orbitals.\cite{Goodenough1958An, Goodenough1955Theory3, Kanamori1959SuperexchangeOrbitals, Kanamori1963ElectronMetals} For tensile strain on STO, the $x^2-y^2$ orbital ordering observed by XLD is predicted to result in AF superexchange interactions between the unoccupied Mn $3z^2-r^2$ and Cr $3z^2-r^2$ orbitals through the apical O \textit{2p} orbital consistent with Figure \ref{fig:XMCD} for the [2 LSCO/ 2 LSMO]$_{10}$ SL on STO. Compressive strain on LAO induces $3z^2-r^2$ orbital ordering which should favor FM ordering between the half filled Mn$^{3+}$ $3z^2-r^2$ and the empty Cr $3z^2-r^2$ orbitals and AF ordering between the empty Mn$^{4+}$ and Cr$^{3+}$ and Cr$^{4+}$ $3z^2-r^2$ orbitals. The observation of anti-aligned Mn and Cr spins from the Mn and Cr L-edge XMCD spectra for the SLs on LAO suggest that the AF superexchange interactions between Mn$^{4+}$ and Cr$^{3+}$ and Cr$^{4+}$ across the interface dominate the spin interactions.  

The lowering of the  energies of the Mn and Cr $z^2-r^2$ orbitals which points out-of-plane for compressive strain on LAO facilitates charge transfer from the LSMO layer to the LSCO layer via the bridging apical O. This is consistent with charge transfer from Mn to Cr as the strain is changed from STO to LAO.

As the LSMO thickness is increased, Mn-Mn exchange interactions in LSMO layers away from the interface become important and the effect on the total magnetization is evident in the [2 LSCO/ 4 LSMO]$_{6}$ SLs. On LSAT, the LSMO is close to cubic and bulk-like FM double-exchange interactions lead to an increase in the total magnetization with LSMO thickness.\cite{colizzi2008interplay, fang2000phase} On STO, tensile strain leads to $x^2-y^2$ orbital ordering and competition between A-type AF and FM leading to a slight reduction in the total magnetization and the Curie temperature compared to LSAT. 
For the 2\% compressive strained samples on LAO with LSMO thicknesses greater than 2 uc, C-type AF ordering \cite{colizzi2008interplay,fang2000phase} competes with FM leading to a lower net magnetic moments compared to the analogous samples on LSAT and STO (Figure \ref{fig:SQUID}(c)). The lower moments may be attributed to a canted AF state.\cite{Moon2017StructuralHeterostructures, grutter2018strain}


A possible explanation for the weak effect of strain on magnetism for the [2 LSCO/ 2 LSMO]$_{10}$ samples may be interfacial Cr/Mn intermixing leading to a solid solution of the LSCO and LSMO layers. However, atomic-resolved EELS results in Figure \ref{fig:TEM} evidence distinct LSCO and LSMO layers. 

For single layer LSMO films, magnetic dead layers are associated with structural, chemical and electronic reconstructions associated with the polar stacking of planes with the LSMO films.\cite{Huijben2008CriticalFilms, Peng2014TuningEngineering, Koohfar2017StructuralInterfaces, Koohfar2019ConfinementHeterostructures} Changes to the LSMO octahedra as a result of interfacial structural coupling may also lead changes to the electronic bandwidth of the LSMO films resulting in magnetically dead layers.\cite{Moon2017StructuralHeterostructures} In the current work, we find that nominally valence-matched and iso-structural LSCO layers are effective in alleviating the interfacial structural and polar mismatch at the LSMO interface leading to the stabilization of a strain-independent FM ground state. Further studies will focus on the effect of the LSCO spacers on the transport properties of ultra-thin LSMO films.

\section{Conclusion}
In conclusion we compare the magnetic properties of [LSCO/LSMO] superlattices as a function of epitaxial strain. The presence of the LSCO layers leads to FM LSMO ground states for both compressive and tensile strained heterostructures, indicating that the removal of magnetic dead layer effects for LSMO is independent of biaxial strain. For [2 LSCO/ 2 LSMO]$_{10}$ SLs, the magnetization is independent of strain and is determined by AF exchange interactions between Cr and Mn. As the LSMO thickness is increased, orbital ordering related to strain results in competing Mn-Mn AF interactions resulting in reduced magnetization for SLs on LAO compared to SLs on STO and LSAT indicative of a relationship between the magnitude of the substrate induced strain via Jahn-Teller distortions of the oxygen octahedra. The stabilization of FM in  ultra-thin 2 uc thick LSMO layers in [LSCO/LSMO] heterostructures allows for the separation of interfacial and bulk contributions to magnetic ordering with important implications for strain engineering in strongly correlated systems.

\section{Acknowledgements}
D.P.K. and S.K. acknowledge financial support by the US National Science Foundation under Grant No.  NSF DMR1751455. This research used resources of the Advanced Light Source, which is a DOE Office of Science User Facility under contract no. DE-AC02-05CH11231.  Use of the Advanced Photon Source was supported by the U.S. Department of Energy, Office of Science, Office of Basic Energy Sciences, under Contract No. DE-AC02-06CH11357. The authors acknowledge use of the SQUID and PPMS facility in the Department of Materials Science and Engineering at North Carolina State University. A portion of this research was conducted at the Center for Nanophase Materials Sciences, which is a DOE Office of Science User Facility. The authors acknowledge the use of facilities within the Eyring Materials Center at Arizona State University. A.B.G. acknowledges discussions with A.J. Millis and support from the Flatiron Institute’s Scientific Computing Core.

%
\end{document}